\documentstyle[12pt]{article}
\begin{document}
\title{QUANTIZED SPACE-TIME AND TIME'S ARROW}
\author{B.G. Sidharth\\ Centre for Applicable Mathematics \& Computer Sciences\\
B.M. Birla Science Centre, Hyderabad 500 063}
\date{}
\maketitle
\footnotetext{Email:birlasc@hd1.vsnl.net.in}
\begin{abstract}
Motivated by the latest direct detection of time assymetry in Kaon decay at
CERN and Fermilab, we suggest a theoretical rationale for this puzzle, in
terms of quantized time.
\end{abstract}
The arrow of time has been a puzzle for a long time. As is well known, the
laws of Newtonian Mechanics, Electromagnetism or Quantum Theory do not
provide an arrow of time - they are equally valid under time reversal, with
only one exception. This is in the well known problem of Kaon decay. On the
other hand it is in Thermodynamics and Cosmology that we find an arrow of
time.\\
It is also true that there has been no theoretical rationale for the Kaon
puzzle which we will touch upon shortly. We will try to find such a
theoretical understanding in the context of quantized space-time, $\sim \hbar /
(\mbox{energy})$, that is the Compton time\cite{r1}. It may be mentioned
that the concept of discrete time or chronon has a long history\cite{r2,r3,r4}.\\
Let us start with one of the simplest quantum mechanical systems, one which can
be in either of two sates separated by a small energy\cite{r5,r6}
\begin{equation}
\imath \hbar \frac{d\psi_\imath}{dt} \approx \imath \hbar [\frac{\psi_\imath
(t+n\tau)-\psi_\imath (t)}{n\tau}] = \sum^{2}_{\imath =1} H_{\imath j}
\psi_\imath\label{e1}
\end{equation}
$$\psi_\imath  = e^{\frac{\imath}{\hbar}Et} \phi_\imath$$
where,\\
$H_{11} = H_{22}$ (which we set $= 0$ as only relative energies of the
two levels are being considered) and $H_{12} = H_{21} =E$,
by symmetry. Unlike in the usual theory where $\delta t = n \tau \to 0$, in the
case of quantized space-time $n$ is a positive integer. So the second
term of (\ref{e1}) reduces to
$$[E + \imath \frac{E^2 \tau}{\hbar}] \psi_\imath = [E (1+\imath )]\psi_\imath,
\mbox{as} \quad \tau = \hbar/E$$
The fact that the real and imaginary parts are of the same order is infact
borne out by experiment.\\
From (\ref{e1}) we see that the Hamiltonian is not Hermitian that is it
admits complex Eigen values indicative of decay, if the life times of
the states are $\sim \tau$.\\
In general this would imply the exotic fact that if a state starts out as
$\psi_1$ and decays, then there would be a non zero probability of seeing
in addition the decay products of the state $\psi_2$. In the process it is
possible that some symmetries which are preserved in the decay of $\psi_1$
or $\psi_2$ separately, are voilated.\\
In this context we will now consider the Kaon puzzle. As is well known from
the original work of Gellmann and Pais, the two state analysis above is
applicable here\cite{r7,r8}. In the words of Penrose\cite{r9}, "the tiny fact of an
almost completely hidden time-asymmetry seems genuinely to be present in
the $K^0$-decay. It is hard to believe that nature is not, so to speak,
trying to tell something through the results of this delicate and beautiful
experiment." On the other hand as Feynman put it\cite{r6}, "if there is
any place where we have a chance to test the main principles of quantum
mechanics in the purest way.....this is it."\\
What happens in this well
known problem is, that given $CP$ invariance, a beam of $K^0$ masons can
be considered to be in a two state system as above, one being the short lived component
$K^S$ which decays into two pions and the other being the long lived state
$K^L$ which decays into three pions. In this case $E \sim 10^{10} \hbar$\cite{r7},
so that $\tau \sim 10^{-10}sec$. After a lapse of time greater than
the typical decay period, no two pion decays should be seen in a beam consisting
initially of the $K^0$ particle. Otherwise there would be violation of
$CP$ invariance and therefore also $T$ invariance. However exactly this
violation was observed as early as 1964\cite{r10}. This violation of time
reversal has now been confirmed directly by experiments at
Fermilab and CERN\cite{r11}.\\
We would like to point out that the Kaon puzzle has a natural explanation
in the quantized time scenario discussed above.

\end{document}